\documentclass[a4paper,11pt]{amsart}

\usepackage[english]{my-shortcuts}
\usepackage{srcltx,algorithm,algorithmic}

\usepackage{hyperref}

\usepackage{graphicx}   
\usepackage{grffile}

\title[Choosing $\lambda$ for basis pursuit]{Hedging parameter selection for basis pursuit}
\author{St\'ephane Chr\'etien, Alex Gibberd  and Sandipan Roy} 

\thanks{S.C. is with the National Physical Laboratory, Hampton Road Teddington, TW11 0LW, UK. Email: stephane.chretien@npl.co.uk}

\thanks{A. G. is with the Department of Mathematics, Imperial College London, London, SW7 2AZ, UK. Email: a.gibberd@imperial.ac.uk}

\thanks{S. R. is with the Department of Statistical Science
University College London, 
Gower Street, London WC1E 6BT,
UK. Email: sandipan.roy@ucl.ac.uk}

\begin{document}
\begin{abstract}
	In Compressed Sensing and high dimensional estimation, signal recovery often relies on sparsity assumptions and estimation is performed via $\ell_1$-penalized least-squares optimization, a.k.a. LASSO. The $\ell_1$ penalisation is usually controlled by a weight, also called "relaxation parameter", denoted by $\lambda$. It is commonly thought that the practical efficiency of the LASSO for prediction crucially relies on accurate selection of $\lambda$. In this short note, we propose to consider the hyper-parameter selection problem from a new perspective which combines the Hedge online learning method by Freund and Shapire, with the stochastic Frank-Wolfe method for the LASSO. Using the Hedge algorithm, we show that a our simple selection rule can achieve prediction results comparable to Cross Validation at a potentially much lower computational cost. 
	\end{abstract}

	\maketitle
	
	{\bf Keywords:} LASSO, Sparse regression, Compressed Sensing, prediction, linear model, hyper-parameter selection.
	
	\bigskip
	
	\section{Introduction}
	
	\subsection{Problem statement and main results}
	Many problems in signal processing can be expressed as the one of recovering a vector $\beta$ from linear measurements of the form
	\begin{align*}
	y  =  X\beta+z
	\end{align*}
	where $X$ denotes a $n\times p$ design matrix, $\beta\in\R^{p}$ is an unknown parameter and the components of the error $z$ are assumed random and sometimes i.i.d. with normal distribution 
	$\mathcal N(0,\sigma^2)$. 
	
	The case where $p$ is much larger than $n$ has been the subject of an extensive research which began with the discovery of Compressed Sensing after the breakthrough papers \cite{candes2006robust} and \cite{donoho2006compressed}.
	One of the main discoveries of Compressed Sensing is that if $\beta$ is sufficiently sparse, one can design matrices $X$ such that the main features of the vector $\beta$ could be recovered by solving the penalized least squares problem 
	\bea
	\label{lasso}
	\qquad \ch{\beta}_\lb & \in & \down{b\in \mathbb R^p}{\rm argmin} \ \frac12\|y-X b\|_2^2+\lb \|b\|_1.
	\eea
	In the Signal Processing community, this approach is called Basis Pursuit.
	In the field of statistics, the solution of this problem is called the LASSO estimator of $\beta$. The acronym LASSO, due to \cite{Tibshirani:JRSSB96}, stems for Least Absolute Shrinkage and Selection Operator. The use of the $\ell_1$-norm penalty has a long and interesting history \cite{Candes:ActaNum06}. One of its main properties is that the components of the standard least-squares 
	estimator $\ch{\beta}$ are shrinked. Most of them are even shrinked to the point of being set to zero. In Signal Processing, this approach can provide impressive results in MRI reconstruction \cite{lustig2008compressed}, gene selection in cancer studies \cite{Tibshirani:JRSSB96}, time series filtering \cite{kim2009ell_1}, etc. One of the main tasks of the statistician is to show that the discovered nonzero components are good predictors for the experiment under study.
	We refer the interested 
	reader to \cite{BuhlmannAndvandeGeer:Springer11} for an overview of the relationships between sparsity and statistics, and 
	sparsity promoting penalizations of the least-squares criterion. 
	
	\subsection{Algorithms}
	A very efficient algorithm, based on Nesterov's method, for 
	solving the LASSO estimation problem is described in \cite{Becker:SIAMImSc10}. Another approach is the one based on Bregman's iterations \cite{yin2008bregman}. 
	A closely related approach is the Alternating Direction Method of Multipliers \cite{parikh2014proximal}. For large scale problems, it is often advised to resort to the Frank-Wolfe algorithm \cite{jaggi2013revisiting}. Research in devising efficient algorithms for solving the LASSO optimisation problem is still a very active trend; 
	see e.g. \cite{tibshirani2017dykstra}, \cite{ndiaye2017efficient}, \cite{massias2018dual}, etc.
	
	Online algorithms for the Basis Pursuit/LASSO have been devised only very recently and very efficient online methods are now available, \cite{wang2013online},  \cite{LafondWaiMoulines:ArXiv2015}.  
	Our approach in the present paper relies in an essential way on such online algorithms. We make the choice of specialising our approach by using the algorithm in \cite{LafondWaiMoulines:ArXiv2015} for the sake of clarity in the exposition. However, any online algorithm for solving the LASSO problem could be put to work in exactly the same way.

    \subsection{Previous work}
    
    The problem of selecting the hyper parameter $\lambda$ being a key step in the estimation procedure, much work has been devoted to it. 
    
    A first approach to choosing $\lambda$ is to take the theoretical value proposed e.g. by \cite{candes2009near}; see also \cite{chretien2014sparse} for the case where the noise variance is unknown. However, theoretical values may depend on unrealistic coefficients coming from crude majorisations in the proofs, and as such may not be so relevant for the problem at hand. The main data driven approaches which are used in practice for solving the hyperparameter calibration problem for the Basis Pursuit/LASSO are 
    \begin{itemize}
        \item Cross Validation \cite{Tibshirani:JRSSB96}, \cite{arlot2010survey}, 
        \item the standard AIC, BIC criteria combined with the LARS \cite{zou2007degrees}, 
        \item Stability selection,
        \item the SURE procedure \cite{zou2007degrees,dossal2013degrees}. 
        \item Universal Quantile Thresholding \cite{giacobino2015quantile}
    \end{itemize}
    The fastest known method seems to be the Universal Quantile Thresholding (UQT) of \cite{giacobino2015quantile}. The UQT method however needs a preliminary Cross-Validation based estimator of the variance when the variance is unknown prior to the experiment. The AIC and BIC combined with the LARS are also very fast, but they come with no guarantee when the design is does not satisfy some kind of incoherence property, i.e. a bound on the maximum scalar product between to different columns, or some Restricted Isometry property. Both the incoherence and the RI property are related in a weak sense; see 
    \cite{chretien2012invertibility}. 
	
	\section{Hedging parameter selection} 
	
	Our approach to hyperparameter selection is based on a combination of Freund and Shapire's Hedge method \cite{freund1995desicion}, \cite{arora2012multiplicative}, and Lafond, Wai and Moulines' online Frank-Wolfe algorithm \cite{LafondWaiMoulines:ArXiv2015}. 

    \subsection{Main idea}
	Instead of choosing $\lambda$ in \eqref{lasso}, one can equivalently choose $r$ in the constrained least-squares problem 
	\begin{align}
	    \hat \beta_r & \in \textrm{argmin}_{b\in \mathbb R^p} \ 
	    \frac12 \Vert y-Xb\Vert_2^2 
	\end{align}
	subject to 
	\begin{align}
	    \Vert b\Vert_1 & \le r.
	\end{align}
	
	We first select a finite set of values $\mathcal R$ from which the values $r$ will be chosen. This set $\mathcal R$ can be chosen exactly in the same way as for Cross Validation. Each value of $r$ can be interpreted as an "expert" and these experts can be compared based on how well they predict the value of $y_i$ given the value of $x_i\in \mathbb R^p$. For this purpose, one can use the Hedge method of Freund and Shapire \cite{freund1995desicion}. In order to apply the Hedge algorithm in the context of hyperparameter selection for the LASSO, one needs to run an online algorithm, which is updated at each step, based on a new observation, and which can be used to predict the next observation. The prediction of the next observation at each step of the online LASSO method, can then be used in order to compute a loss that can itself be incorporated into the update step of Freund and Shapire's method.
	
	More precisely, our method simply consists in running several Stochastic 
	Franck-Wolfe algorithms in parallel, each one with a different candidate value of the parameter $r\in \mathcal R$. At each step of the Stochastic Franck-Wolfe algorithm, the two following steps are taken:
	
	\begin{itemize}
	    \item a prediction is performed for the next observed value and an error $err_r^{(i)}$ is measured for every $r\in \mathcal R$, 
	    \item a loss $loss_r^{(i)}$ is computed for each value $r\in \mathcal R$, 
	    \item a probability, denoted by 
	$h_r^{(i)}$, is associated to each $r\in \mathbb R$ and the probability vector $h^{(i)}$ is updated according to the rule 
	\begin{align}
        h_r^{(i+1)} & = \frac{h_r^{(i)} \ \exp \left(-\beta \ loss_r^{(i)}\right)}{\sum_{r\in \mathcal R} \ h_r^{(i)} \ \exp \left(-\beta \ loss_r^{(i)}\right)}.
	\end{align}
\end{itemize}

    	\subsection{The algorithm}
    	Our method is presented in Algorithm \ref{HedgeStochFW} below. 
	Since we have $n$ observations, the 
	algorithm stops after $n$ steps. The Hedge vector $h^{(n)}$ obtained as an output of the algorithm is a probability vector which can be used for the purpose of predictor aggregation. In the 
	case where the output Hedge vector is a Dirac vector, up to numerical tolerance, the Hedge algorithm 
	selects only one predictor. 
	
	\begin{algorithm}
		\caption{HedgeStochFW for selection of $r$ in the LASSO \label{HedgeStochFW}}
		\begin{algorithmic} 
			\REQUIRE $A \in \mathbb R^{n\times p}$, $y \in \mathbb R^n$, $\eta>0$ and a set $\mathcal R$ of 
			candidate values for $r$.
			\STATE Initialize with $h=1$, $b_r^{(0)}=0$, $r\in \mathcal R$, $\overline{\beta}_i=0$ and $\overline{\alpha}_i=0$.
			\FOR{$i = 1,\ldots,n$}
			\FOR{$r \in \mathcal R$}
			\STATE Compute the squared error and the pre-hedge vector 
			\begin{align*}
			\hat{\epsilon}_r^{(i)} & = ( y_i-A_i b_r^{(i-1)} )^2  
			\end{align*} 
			and 
			\begin{align*}
			\tilde{h}_r^{(i)} & = h_r^{(i-1)} \exp \left(-\eta \ \hat{\epsilon}_r^{(i)}\right).
			\end{align*}
			\STATE Evaluate 
			\begin{align}
			\overline{\beta}_r^{(i)} & = \Big(1-\frac{1}{i}\Big) \ \overline{\beta}_r^{(i-1)} +\frac{1}{i} \ A_i^ty_i \\
			\overline{\alpha}_r^{(i)} & = \Big(1-\frac{1}{i}\Big) \ \overline{\alpha}_r^{(i-1)} +\frac{1}{i} \ A_i^tA_i
			\label{alph}
			\end{align}
			\STATE Form the gradient estimate 
			\begin{align}
			\nabla f (b_r^{(i-1)}) & = \overline{\alpha}_r^{(i)} \ b_r^{(i-1)} - \overline{\beta}_r^{(i)}.
			\label{grad}
			\end{align}
			\STATE Find 
			\begin{align*}
			j_{\max} & \in {\rm argmax}_{j=1}^p \ \left| \left(\nabla f(b_r^{(i-1)})\right)_{j}\right|
			\end{align*}
			\STATE Set 
			\begin{align*}
			d_r^{(i)} & = - r \ {\rm sign} \ \left(\nabla f(b_r^{(i-1)})\right)_{j_{\max}}\ e_{j_{\max}}
			\end{align*}
			\STATE Update 
			\begin{align*}
			b_r^{(i)} & = \left(1-\frac{K}{t+K-1}\right) \ b_r^{(i-1)} + \frac{K}{t+K-1} d_r^{(i)}.
			\end{align*}
			\ENDFOR
			\STATE Update 
			\begin{align*}
			h^{(i)} & = \frac1{\sum_{r\in \mathcal R }\ \tilde{h}_r^{(i)}} \ \tilde{h}^{(i)}.
			\end{align*}
			\ENDFOR
			\STATE Output $h^{(n)}$ and $b_r^{(n)}$, $r\in \mathcal R$.
		\end{algorithmic}
		\label{alg}
	\end{algorithm}

	\section{Numerical study} 
    \subsection{Comparison with Cross Validation: Gaussian i.i.d. design}	
	We performed several simulation with the LASSO for sparse signal recovery. In these experiments, the number of observations was taken as $80$ and $100$, and the dimension of the sparse signal was  $200$. The sparsity was set to $s_0=5$. The non-zero components' location were chosen uniformly at random and their values were drawn independently at random as the product of a $\pm 1$ Bernoulli variable times $1+G$, where $G$ is a standard Gaussian $\mathcal N(0,1)$ variable. The standard deviation of the observation noise was taken as $.1$, $.01$ and $.001$.

\begin{figure}[!ht]
        \includegraphics[width=12cm]{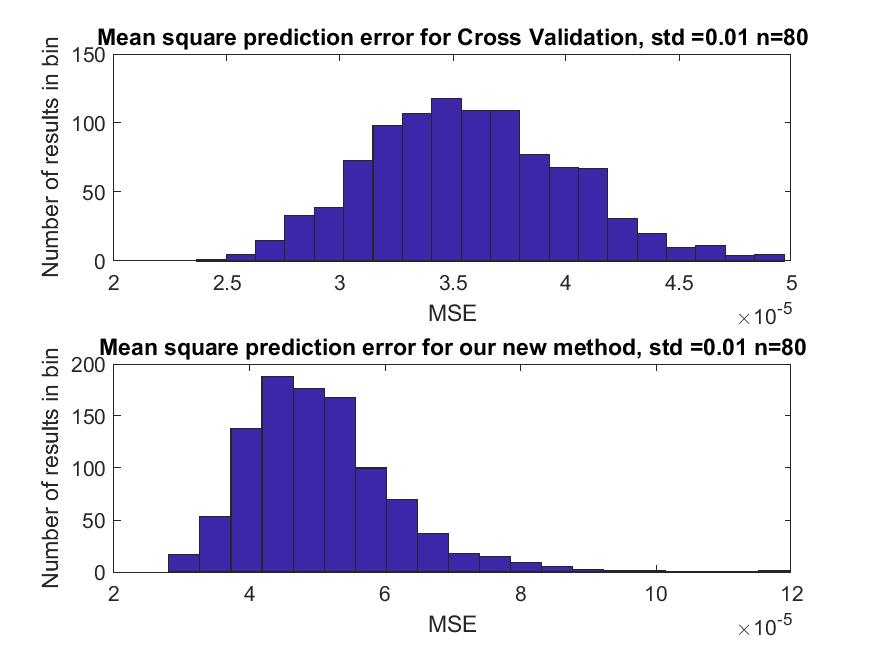}
		\caption{The top figure shows the $1/\sqrt{n}$ normalised least squares prediction error for the Cross Validation approach and the bottom figure shows the same prediction error for our method based on 1000 Monte Carlo experiments.}
        \includegraphics[width=12cm]{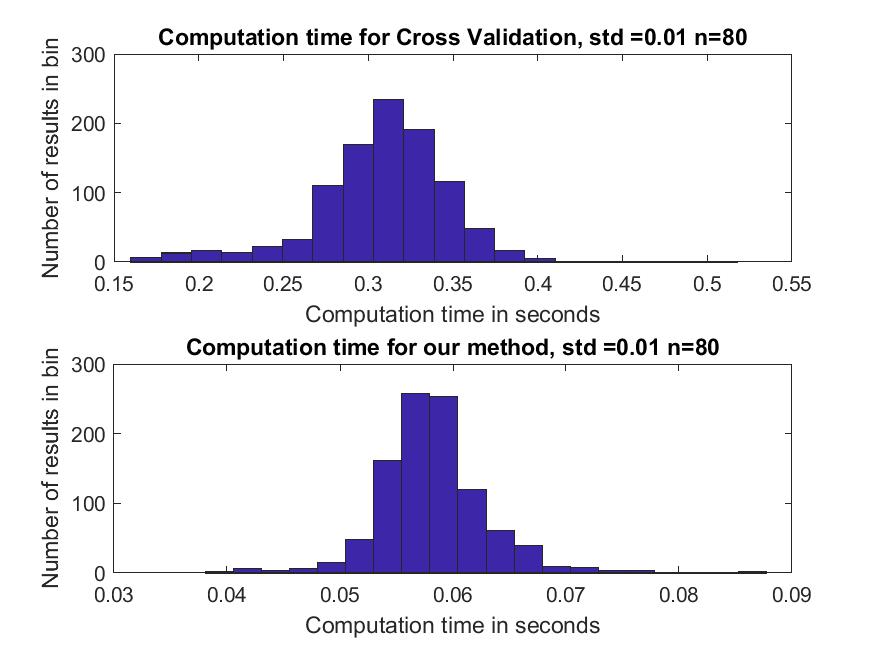}
		\caption{The top figure shows the computation time for the Cross Validation approach and the bottom figure shows the computation time for our method for the same experiments based on 1000 Monte Carlo trials.}
		\label{1080}
	\end{figure}

\begin{figure}[!ht]
        \includegraphics[width=12cm]{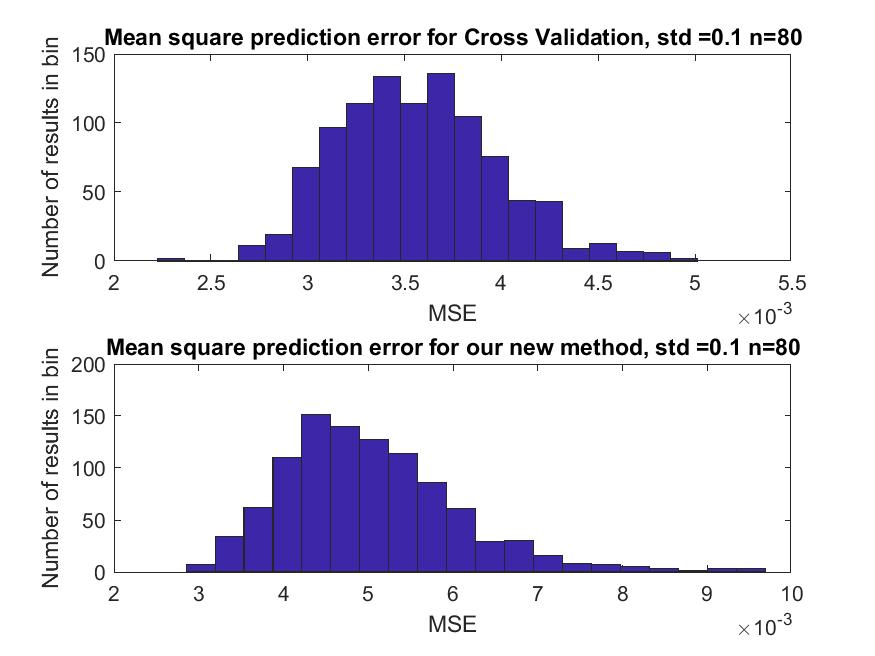}
		\caption{The top figure shows the $1/\sqrt{n}$ normalised least squares prediction error for the Cross Validation approach and the bottom figure shows the same prediction error for our method based on 1000 Monte Carlo experiments.}
        \includegraphics[width=12cm]{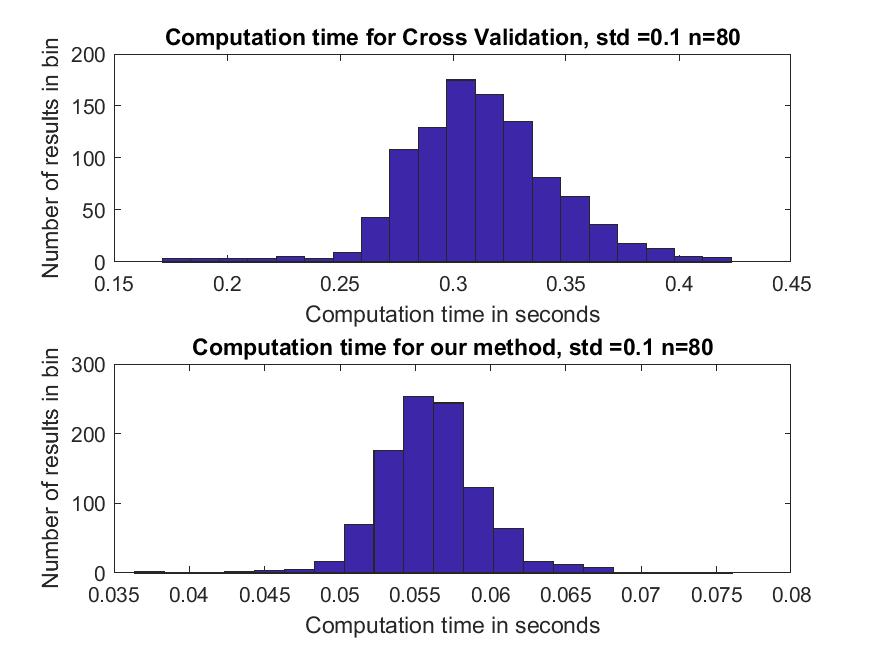}
		\caption{The top figure shows the computation time for the Cross Validation approach and the bottom figure shows the computation time for our method for the same experiments based on 1000 Monte Carlo trials.}
		\label{10080}
	\end{figure}

\begin{figure}[!ht]
        \includegraphics[width=12cm]{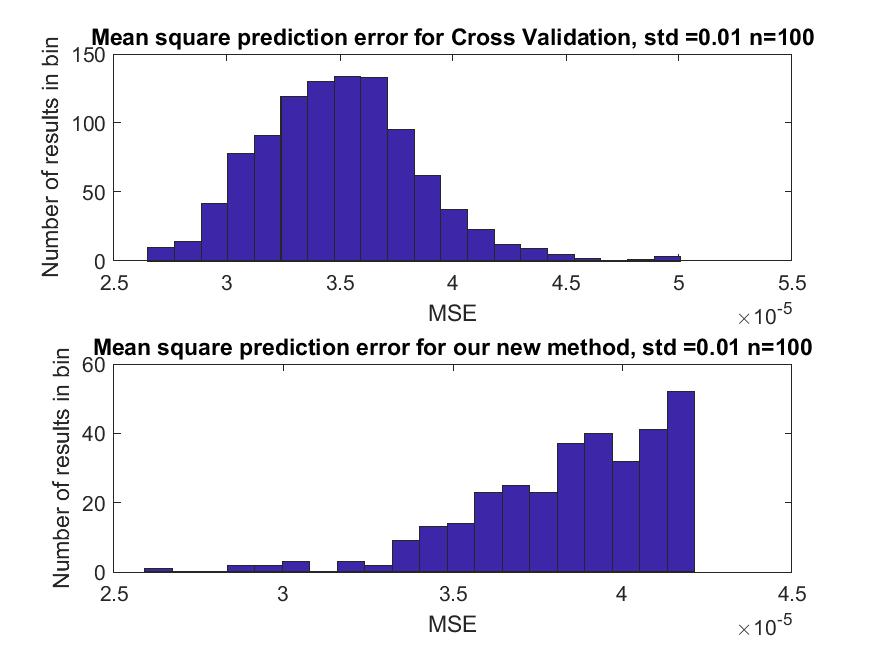}
		\caption{The top figure shows the $1/\sqrt{n}$ normalised least squares prediction error for the Cross Validation approach and the bottom figure shows the same prediction error for our method based on 1000 Monte Carlo experiments.}
        \includegraphics[width=12cm]{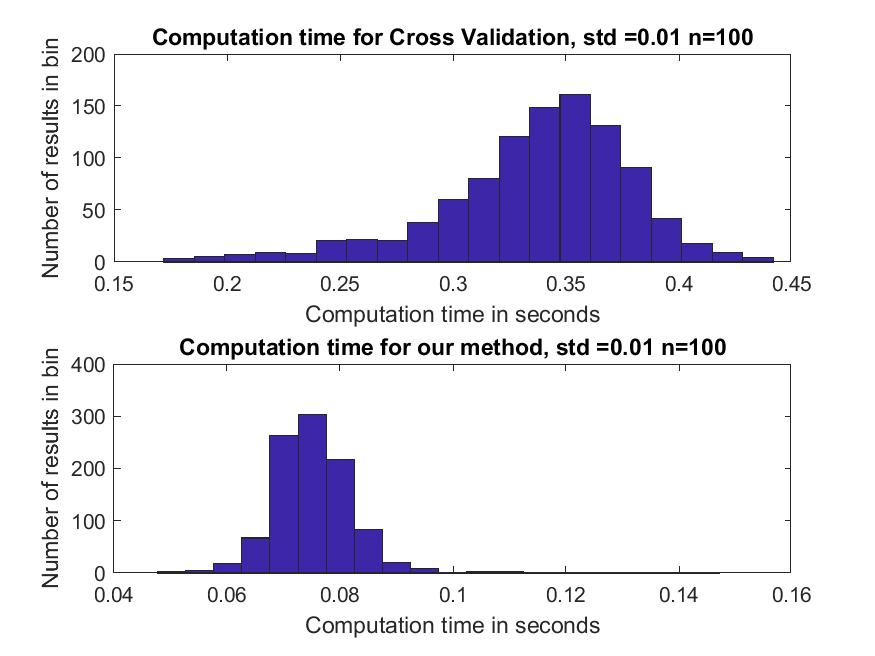}
		\caption{The top figure shows the computation time for the Cross Validation approach and the bottom figure shows the computation time for our method for the same experiments based on 1000 Monte Carlo trials.}
		\label{10100}
	\end{figure}

\begin{figure}[!ht]
        \includegraphics[width=12cm]{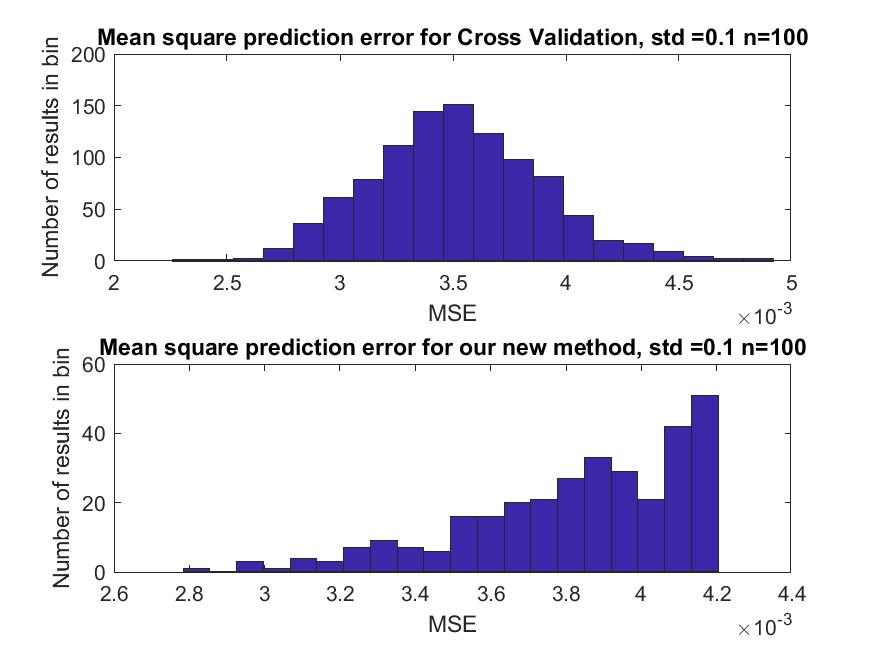}
		\caption{The top figure shows the $1/\sqrt{n}$ normalised least squares prediction error for the Cross Validation approach and the bottom figure shows the same prediction error for our method based on 1000 Monte Carlo experiments.}
        \includegraphics[width=12cm]{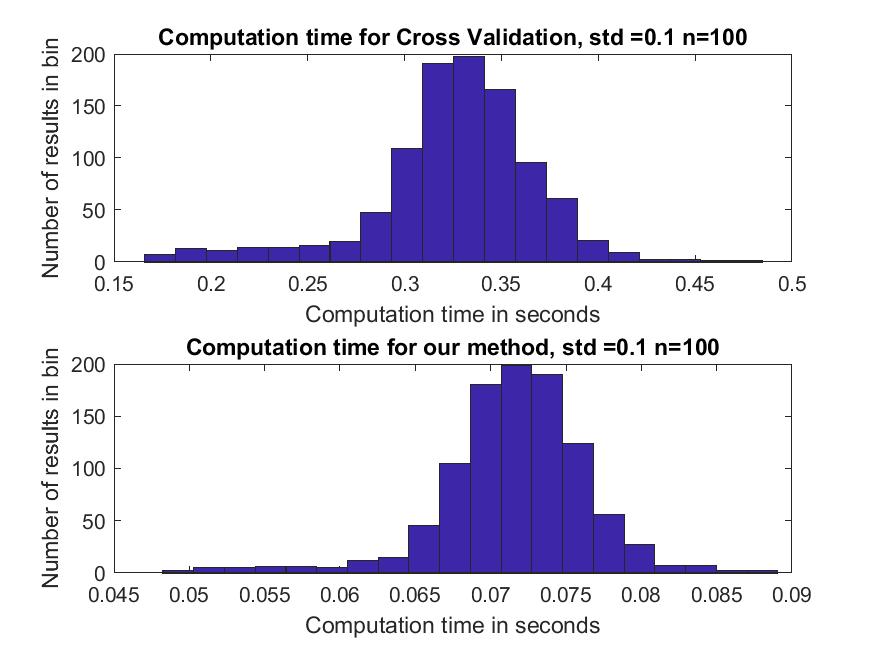}
		\caption{The top figure shows the computation time for the Cross Validation approach and the bottom figure shows the computation time for our method for the same experiments based on 1000 Monte Carlo trials.}
		\label{100100}
	\end{figure}

Figures \ref{1080}, \ref{10080}, \ref{10100} and 
\ref{100100} show the result of 1000 Monte Carlo experiments. The performance of the method is compared with the ones of the 5-fold Cross-Validation procedure implemented in the Matlab function {\em lasso}. The figures show comparison of the mean-squared error for the 5-fold Cross-Validation procedure and the estimator provided by Algorithm \ref{alg} and shows that the mean squared error is as good for Algorithm \ref{alg} as for Matlab's Cross Validation procedure. The figures also display the computation times associated with the two methods.       

The experiments show that the same order of MSE can be obtained for both method while the new proposed Hedge based LASSO is one order of magnitude faster.

	
	\subsection{Comparison with Cross Validation: A gene expression design}
	We also performed several experiments with the LASSO based on a highly correlated design previously studied in \cite{chretien2015investigating} and \cite{chretien2016bregman}. The columns of the design matrix are the expression of 34 genes for 100 patients. 
	
\begin{figure}[!ht]
        \includegraphics[width=12cm]{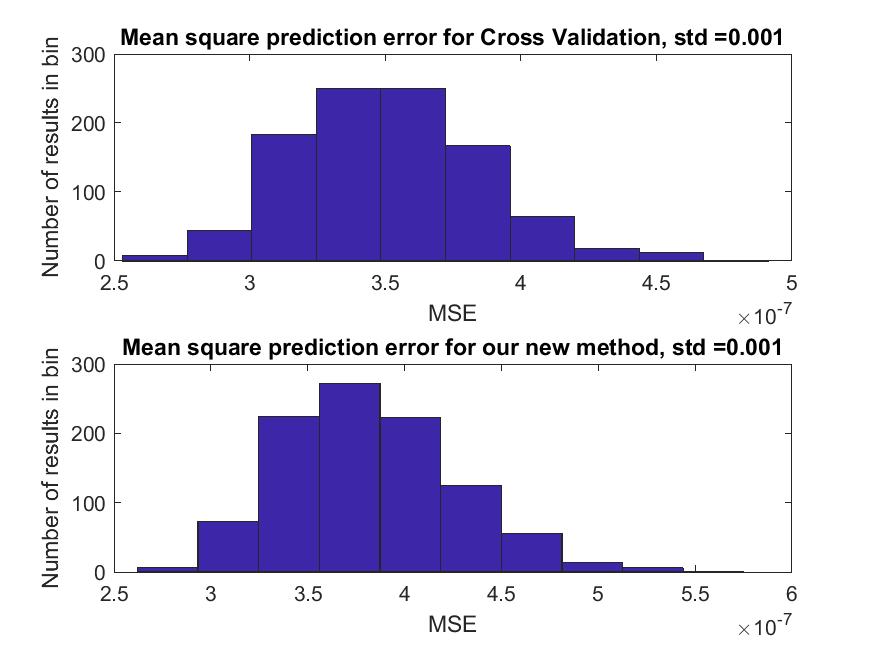}
		\caption{The top figure shows the $1/\sqrt{n}$ normalised least squares prediction error for the Cross Validation approach and the bottom figure shows the same prediction error for our method based on 1000 Monte Carlo experiments.}
        \includegraphics[width=12cm]{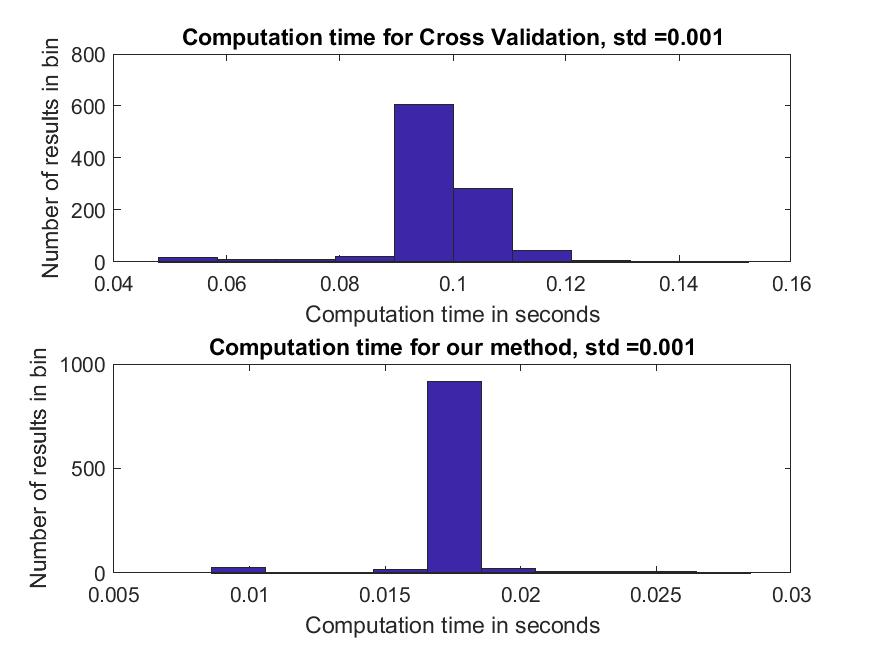}
		\caption{The top figure shows the computation time for the Cross Validation approach and the bottom figure shows the computation time for our method for the same experiments based on 1000 Monte Carlo trials.}
		\label{gene1}
	\end{figure}

\begin{figure}[!ht]
        \includegraphics[width=12cm]{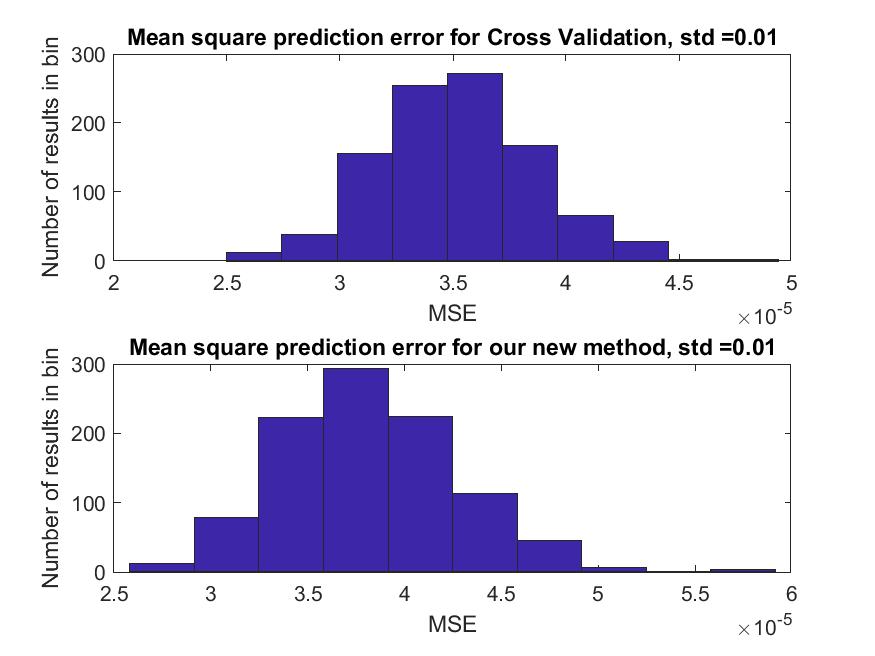}
		\caption{The top figure shows the $1/\sqrt{n}$ normalised least squares prediction error for the Cross Validation approach and the bottom figure shows the same prediction error for our method based on 1000 Monte Carlo experiments.}
        \includegraphics[width=12cm]{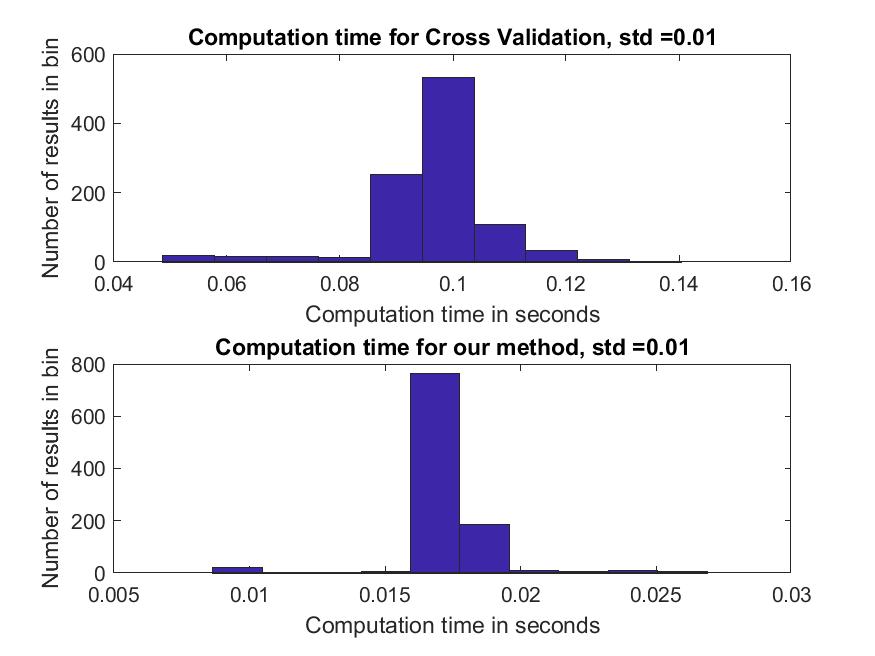}
		\caption{The top figure shows the computation time for the Cross Validation approach and the bottom figure shows the computation time for our method for the same experiments based on 1000 Monte Carlo trials.}
		\label{gene2}
	\end{figure}

\begin{figure}[!ht]
        \includegraphics[width=12cm]{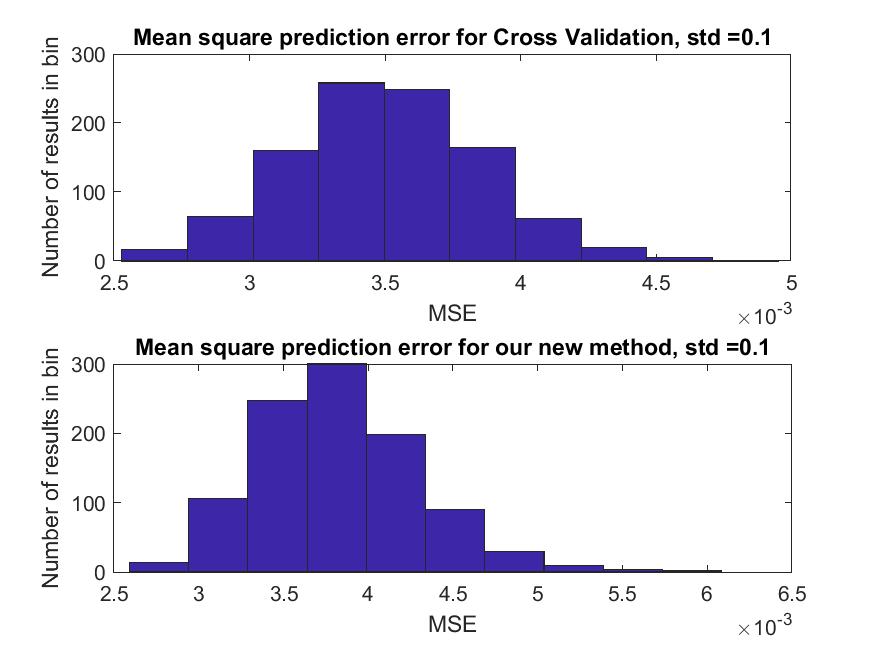}
		\caption{The top figure shows the $1/\sqrt{n}$ normalised least squares prediction error for the Cross Validation approach and the bottom figure shows the same prediction error for our method based on 1000 Monte Carlo experiments.}
        \includegraphics[width=12cm]{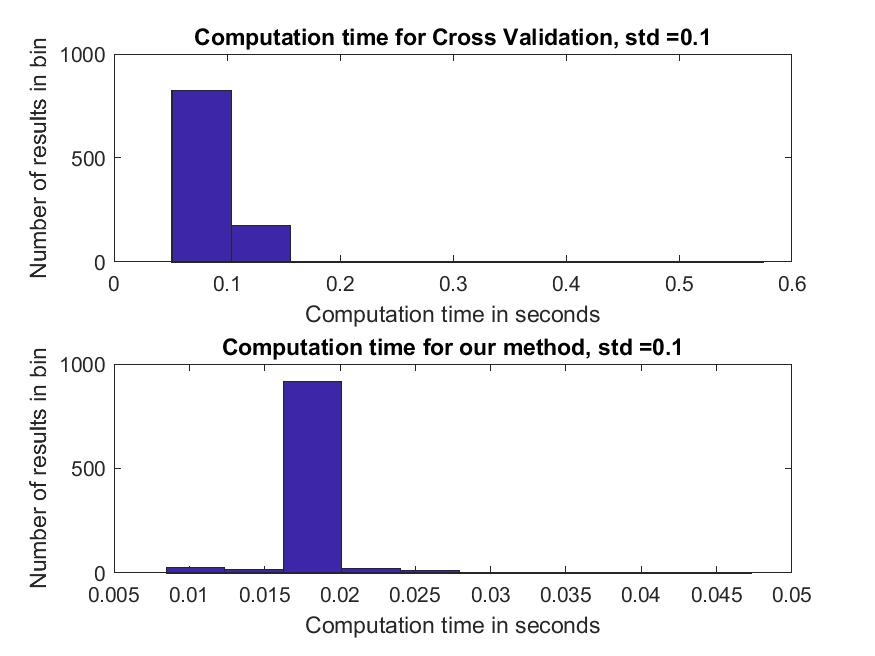}
		\caption{The top figure shows the computation time for the Cross Validation approach and the bottom figure shows the computation time for our method for the same experiments based on 1000 Monte Carlo trials.}
		\label{gene3}
	\end{figure}

	Figure \ref{gene1}, \ref{gene2} and \ref{gene3} show the comparison results for the MSE and computational times for the Cross Validation procedure and our new method. Figure \ref{gene1} (resp. Figure \ref{gene2}, Figure \ref{gene3}) presents the results for a noise level $\sigma$ equal to .1, (resp. .01, .001). As in the case of a random design, our approach is much faster than the optimised routine available in Matlab, despite the fact that our implementation uses quite basic Matlab coding. 
	
	\section{Conclusion and future work}
	
	This short note presents a simple to implement method for choosing the hyper-parameter in the LASSO estimator. Application of this method can also easily be extended to various other models, such as two-stage estimation \cite{chretien2010alternating}, \cite{zhang2010analysis}, generalised linear models \cite{van2008high}, graphical models \cite{meinshausen2006high}, clustering \cite{hocking2011clusterpath}, Robust PCA \cite{candes2011robust}, etc. 
	
	One of the interesting features of the method is that it is potentially robust with respect to dependencies between the observations and event adversarial noise; see \cite{cesa2006prediction}. More results in this direction as well as a theoretical analysis the method will be included in future developments of this preliminary study. 
	
	\bibliographystyle{amsplain}
	\bibliography{LASSOLambdaSelect}

\end{document}